\documentclass[prd,twocolumn,showpacs,preprintnumbers,floats,floatfix,
amssymb,amsmath,nofootinbib]{revtex4}
\usepackage{graphicx}
\usepackage{dcolumn}
\usepackage{bm}
\def\spose#1{\hbox to 0pt{#1\hss}}
\def\lta{\mathrel{\spose{\lower 3pt\hbox{$\mathchar"218$}}
     \raise 2.0pt\hbox{$\mathchar"13C$}}}
\def\gta{\mathrel{\spose{\lower 3pt\hbox{$\mathchar"218$}}
     \raise 2.0pt\hbox{$\mathchar"13E$}}}
\newcommand{\be}{\begin{equation}}
\newcommand{\en}{\end{equation}}
\newcommand{\bea}{\begin{eqnarray}}
\newcommand{\ena}{\end{eqnarray}}

\begin{document}

\title{Trans-Planckian Physics and the Spectrum of Fluctuations
in a Bouncing Universe}

\author{Robert~H.~Brandenberger\footnote{Address from 9/15/2001 -
3/15/2002: Theory Division, CERN, CH-1211 Geneva 23, Switzerland.} and
Sergio~E.~Jor\'as\footnote{Current address: Instituto de F\'{\i}sica, 
Universidade Fe\-de\-ral do Rio de Janeiro, Caixa Postal 68528, Rio de 
Janeiro, RJ 21945-970, Brazil.}} 
\email{rhb@het.brown.edu, joras@het.brown.edu}
\affiliation{Department of Physics, Brown University, Providence, RI
02912, USA} 
\author{J\'er\^ome Martin} 
\email{jmartin@iap.fr}
\affiliation{Institut d'Astrophysique de Paris, 98bis  boulevard Arago,
75014 Paris, France} 
\date{\today}

\begin{abstract}
In this paper, we calculate the spectrum of 
scalar field fluctuations in a bouncing, asymptotically flat Universe, and 
investigate the dependence of the result
on changes in the physics on length scales shorter than the
Planck length which are introduced via modifications of the dispersion
relation. In this model, there are no ambiguities concerning the
choice of the initial vacuum state. We study an example in
which the final spectrum of fluctuations depends sensitively on
the modifications of the dispersion relation without needing to
invoke complex frequencies. Changes in the amplitude and in the 
spectral index are possible, in addition to modulations of the spectrum.  
This strengthens the conclusions of previous work in which the
spectrum of cosmological perturbations in expanding inflationary 
cosmologies was studied, and it was found that, for dispersion
relations for which the evolution is not adiabatic, the spectrum 
changes from the standard prediction of scale-invariance.
\end{abstract}

\maketitle

\section{Introduction}

The dependence of the spectrum of cosmological fluctuations in an 
inflationary Universe on hidden assumptions about the physics on length 
scales much smaller than the Planck length has been recently studied 
in Refs. \cite{BM1,MB2,N1}. {\it A priori} this dependence comes about since 
in typical scalar-field-driven inflationary models the duration of the 
period of
expansion is so long that the physical wavelengths of comoving modes which
correspond to the present large-scale structure of the Universe are much 
smaller than the Planck length at the beginning of inflation. In weakly
coupled scalar field models of inflation, the spectrum of fluctuations is
calculated by assuming that fluctuation modes start out in the vacuum 
(i.e. minimum energy density) state at the beginning of inflation and 
subsequently evolve as determined by the equations of motion. It is hence
not unreasonable to expect that modifications of the physics on
trans-Planckian scales could affect the final spectrum of fluctuations.
\par
While the correct theory of trans-Planckian physics is not known, possible
effects of the new physics can be modeled by changes in the dispersion
relation of the fields corresponding to linear cosmological fluctuations.
Such an approach was used earlier \cite{Unruh,CJ} to study the possible
dependence of the spectrum of black hole radiation on trans-Planckian
physics. Indeed, for a class of dispersion relations which 
deviate so strongly from the usual one such that the evolution of the
mode functions is highly non-adiabatic and the effective frequency
becomes imaginary, it was found that the spectrum of
fluctuations is modified in a significant way compared to what is obtained
for a linear dispersion relation \cite{BM1,MB2}. As shown in 
\cite{MB3,NP2,Starob}, the final spectrum of fluctuations is 
unchanged if the change in the dispersion relation is not too drastic,
the wave function evolves adiabatically, and hence the WKB approximation is 
valid throughout the evolution.
\par
Two major deficiencies in the previous work concerned the choice of
initial conditions and the fact that complex dispersion relations
were required in order to obtain significant deviations from the
standard results. In the previous  papers, 
the assumption was made that the state starts out 
as the one minimizing the energy density. However, if the dispersion 
relation differs greatly from the linear one, in particular if it becomes
complex, the physical motivation for choosing this state becomes unclear.
This problem does not arise in a bouncing Universe. In this
case, modes corresponding to present cosmological scales could well have 
had a physical wavelength smaller than the Planck length at the bounce
point, but at early (pre-bounce) times, the wavelength was larger than the
Planck length. In this context, initial conditions can be set up at some
very early pre-bounce time during which the mode obeys a linear
dispersion relation and the physical meaning of the minimum energy density
state is clear. It is thus interesting to study the spectrum of
fluctuations in a bouncing Universe which is asymptotically flat such
that the choice of initial vacuum becomes unambiguous. It is 
also interesting to exhibit a case where complex dispersion 
relations are not needed. Although complex frequencies are standard 
in classical physics and in quantum mechanics, they seem to be 
more problematic in the context of quantum field theory.
\par
In this article, we find an example of a bouncing Universe in which
the spectrum of fluctuations depends sensitively on changes in the 
physics at length scales smaller than the Planck length, without
requiring complex frequencies at high wavenumbers, thus improving
on the second major deficiency of previous work. Another reason 
for investigating the dependence of the spectrum of 
fluctuations on modifications of trans-Planckian physics in a bouncing 
Universe is that one could expect
(however, as we show, this is not correct) that the deviations in the spectrum
incurred during the pre-bounce period when the wavelength is 
blue-shifting in the trans-Planckian regime will cancel with the 
deviation in the post-bounce trans-Planckian regime when the mode is
being red-shifted. Such a result would then lead to the expectation that
the results on the dependence of the spectrum of fluctuations on
changes in the trans-Planckian physics might be different in the case
of black holes and (non-bounce) inflationary cosmology. A third 
reason for analyzing a bouncing Universe
is that there has recently been a lot of interest
in the field of string cosmology in bouncing
Universe models. In the Einstein frame, the
pre-big-bang scenario \cite{PBB} starts in a collapsing
dilaton-dominated phase, and the same is true
in the Ekpyrotic scenario \cite{KOST}. Our arguments
show that trans-Planckian effects could change
the predictions of standard cosmological perturbation
theory in these examples \footnote{Note that there
is disagreement on the result of the linear theory
of fluctuations in the Ekpyrotic scenario 
\cite{KOST,Lyth,BF,KOST2,Hwang,MPPS}}.
\par
A model for the evolution of the scale factor in a bouncing inflationary
Universe is $a(t)=a_0 {\rm cosh}(H t)$, where $a_0$ denotes the 
value of $a(t)$ at the center of the bounce, and 
$H$ is the Hubble expansion (contraction) rate during the period of 
exponential expansion (contraction). The idea of the calculation is to
assume an initial spectrum of fluctuations at some early time $-t_{\rm f}$
when the wavelength of all modes of interest is much larger than the
Planck length, to propagate the perturbation modes through the bounce
and calculate the final spectrum at time $t_{\rm f}$. In this model, one of
the difficulties encountered in the previous work 
\cite{BM1,MB2}, namely the problem of specifying the
initial state in the trans-Planckian regime, is overcome. However, it
is necessary now to set up initial conditions on wavelengths which
typically also are much larger than the Hubble radius. The latter
problem can be overcome by considering a bouncing Universe which is
radiation-dominated in the asymptotic past and future, e.g. \cite{AS,BD}
$a(\eta)=\sqrt{\alpha ^2 + \beta ^2 \eta^2}$, where $\alpha $ and 
$\beta $ are constants. Another toy model in which it 
is possible to set up well motivated initial
conditions on scales much smaller than the Hubble radius and at the same
time much larger than the Planck scale is a model in which the Universe
is asymptotically flat both at large positive and negative times:
\begin{equation} 
\label{model2}
a(\eta) \, = \, \ell_0 - {{\ell_0 - \ell_{\rm b}} 
\over {1 + (\eta/\eta_0)^2}} \,
\end{equation}
where $\ell_0$ and $\ell_{\rm b}$ (with $\ell_{\rm b} \ll \ell_0$) denote 
the asymptotic size of the Universe and the size at the bounce, 
respectively, and $\eta_0$
determines the time scale of the bounce, i.e. the time over which
the scale factor changes. In the above, we are using conformal time 
$\eta$, and employing the convention that the scale factor carries
dimension of length while the comoving coordinates (including
conformal time) are dimensionless.
\par
Finally, a comment is in order on the issue of backreaction. It has 
been shown in Refs.~\cite{Tanaka,Starob} that the energy density 
of the particles created by trans-Planckian physics can contribute 
in a significant way to the background energy density. We do not 
address this interesting question since this clearly comes as a 
second step once a case where the spectrum is modified has been 
explicitly exhibited. In this paper we concentrate on the first 
question since it would be useless to treat the backreaction 
question if no well-motivated example where the spectrum is 
modified can be found. However, it should be clear that, once this 
is done (as shown in the present article), this problem becomes the 
central question in the study of trans-Planckian physics in 
cosmology.
\par
This article is organized as follows. In section II, we review the 
arguments which show that no change in the final spectrum of
fluctuations of a scalar matter field in a bouncing Universe 
is expected independent of the dispersion relation, provided that
the WKB approximation remains justified. We give a qualitative reason
to expect changes in the amplitude of the final spectrum for
dispersion relations for which the adiabatic approximation breaks
down. As mentioned above, our analysis assumes that
perturbation theory remains justified
and back-reaction can be neglected \cite{Tanaka}. In Section III 
we study a concrete model, namely the
asymptotically flat bouncing Universe given by (\ref{model2})
with a dispersion relation modified in the trans-Planckian
regime according to the prescription of Corley and Jacobson \cite{CJ}.
We show that, as expected, both the overall amplitude and the
shape of the spectrum differ from what is obtained for the linear
dispersion relation, without ever requiring the effective frequency
to become imaginary. The change in the spectrum is produced by an
interesting combination of effects due to the modified dispersion
relation and the driving term for Parker particle production \cite{Parker}. 

\section{Method and Qualitative Considerations}

The equation of motion for a minimally coupled free scalar matter field 
$\Phi(\eta, \bbox{x})$ in a Universe with scale factor $a(\eta)$ in 
momentum space takes on the simple form
\begin{equation} \label{eom1}
\mu ''+ \biggl(n^2 - {{a''} \over a}\biggr) \mu \, = \, 0 \, ,
\end{equation}
making use of the re-scaling
\begin{equation}
\Phi(\eta ,\bbox{x}) \, = \, {1 \over {(2 \pi)^{3/2}}} 
\frac{1}{a(\eta )}\int {\rm d}\bbox{n} 
\mu (n,\eta )e^{i \bbox{n} \cdot \bbox{x}} \, ,
\end{equation}
where $\bbox{n}$ denotes the comoving wavenumber linked to the physical 
wavenumber by the relation $\bbox{k}=\bbox{n}/a(\eta )$.
In the above, a prime denotes the derivative with respect to conformal
time. The previous equations are valid for a spacetime with 
flat spatial sections. The advantage of considering a test 
scalar field and/or gravitational waves is that we do not need 
to address the origin of the dynamics of the scale factor. This 
is in contrast to the case of density perturbations. However, 
the Friedmann equation reads ${\cal H}^2/a^2+K/a^2=\kappa \sum _i\rho _i /3$, 
where ${\cal H}\equiv a'/a$ and $\rho _i>0$ is the energy density. This 
means that a bounce is consistent with the Einstein equations 
only if $K=1$. For $K=0$, it would imply $\rho =0$ for each 
component. Therefore, it seems that it is inconsistent to use 
Eq.~(\ref{eom1}) if the scale factor behaves such that there is 
a bounce. There are two ways out of the previous argument. The 
first one is the following. If $K=1$, then Eq.~(\ref{eom1})
takes the form $\mu ''+(n^2-K-a''/a)\mu =0$. In addition 
the eigenfunctions of the three-dimensional Laplacian operator 
are no longer planes waves. However, if we restrict ourselves 
to modes with wavelengths much smaller than the curvature radius, 
then they do not feel the curvature of the spacelike section and 
we can safely work with Eq.~(\ref{eom1}). It could be 
checked that this is indeed the case for the wavenumbers 
considered in the example presented in the following 
section. Another possibility is 
to use the ``nonsingular Universe construction''~\cite{bms}. In this 
case, it is definitely possible to get a spatially flat
bounce. What happens is that the higher
derivative terms become important at the bounce
and enable the evasion of the previous argument by
supplying other terms in the analog of the
first Friedmann equation. In conclusion, although 
in this article we consider a scale factor with a 
bounce, we can nevertheless work with Eq.~(\ref{eom1}), i.e. we 
let the nonsingular Universe construction enter only via the 
scale factor.
\par
The method introduced in \cite{Unruh,CJ} to study the dependence
of the spectrum of fluctuations on trans-Planckian physics is to
replace the linear dispersion relation $\omega _{_{\bf phys}}=k$ by 
a non standard dispersion relation 
$\omega _{_{\bf phys}}=\omega _{_{\bf phys}}(k)$ where the function 
$\omega _{_{\bf phys}}(k)$ is {\it a priori} arbitrary. In the 
context of cosmology, it has been shown in Ref.~\cite{BM1,MB2} that 
this amounts to replacing $n^2$ appearing in (\ref{eom1}) with 
$n_{\rm eff}^2(n,\eta )$ defined by
\begin{equation}
n^2 \, \rightarrow \, n_{\rm eff}^2(n,\eta ) \equiv 
a^2(\eta )\omega _{_{\bf phys}}^2\biggl[\frac{n}{a(\eta )}\biggr].
\end{equation}
For a fixed comoving mode, this implies that the dispersion relation 
becomes time-dependent. Various forms for $n_{\rm eff}$ have been used, e.g.
\begin{widetext}
\begin{equation} 
\label{disp12}
n_{\rm eff}(n,\eta ) = n {{\lambda(\eta)} \over {\ell_{_{\rm C}}}} 
{\rm tanh}^{1/p} \biggl[\biggl({{\ell_{_{\rm C}}} \over 
{\lambda(\eta)}}\biggr)^p \biggr], \quad 
n_{\rm eff}^2(n,\eta ) = n^2 + n^2 b_m \biggl[{{\ell_{_{\rm C}}} 
\over {\lambda(\eta)}} \biggr]^{2m}, 
\end{equation}
\end{widetext}
the first being a generalization of the relation studied by Unruh \cite{Unruh} 
whereas the second one is a generalization of the relations used by
Corley and Jacobson \cite{CJ}. In 
the above, $\lambda(\eta)$ is the physical wavelength, $\ell_{_{\rm C}}$ is 
the cutoff length (which can be taken to be the Planck length), $p$ and $m$ are
integers, and $b_m$ is a real number which can be either positive or
negative. Thus, the equation of motion to be analyzed is
\begin{equation} 
\label{eom2}
\mu '' + \biggl[n_{\rm eff}^2(n,\eta ) - {{a''} 
\over a}\biggr]\mu \, = \, 0 \, .
\end{equation}

For fixed comoving wavenumber $n$, the evolution of $\mu$ depends
crucially on which region $\eta$ lies in. The first region is 
the trans-Planckian region with 
\begin{equation}
\lambda (\eta ) \, \ll \, \ell _{_{\rm C}} \,\,\, ({\rm{Region \ \ 1}}) \, .
\end{equation} 
Let us first consider the standard inflationary scale factor and a monotonic 
dispersion relation [like the first one in Eq. (\ref{disp12}) or the 
second one with $b_m>0$]. This means that, initially, the term 
$n_{\rm eff}^2(n,\eta )$ dominates in Eq. (\ref{eom2}). The initial 
conditions are fixed in this region and since the WKB approximation 
is applicable, we can choose the initial state as the ``minimizing 
energy state'' \cite{Brown78}. Then, the (positive frequency) solution 
is given by 
\begin{equation} \label{WKBsol}
\mu (\eta )\, \simeq \, \frac{1}{\sqrt{2n_{\rm eff}(n,\eta )}}
\exp\biggl[-i\int _{\eta _{\rm i}}^{\eta }n_{\rm eff}{\rm d}\tau \biggr]
\, ,
\end{equation} 
where $\eta_i$ is some initial time. 

The second region corresponds to 
\begin{equation}
\lambda \, \gg \, \ell _{_{\rm C}} \,\,\, {\rm{and}} \,\,\, 
n^2 \, \gg \, a''/a \,\,\, ({\rm{Region \ \ 2}}) \, . 
\end{equation}
In this region the mode has reached the linear part 
of the dispersion relation. The general solution in
Region 2 is the plane wave 
\begin{equation}
\mu \, = \, B_1\exp(-in\eta )+B_2\exp(in\eta )
\end{equation}
with constant coefficients $B_1$ and $B_2$. For the standard
dispersion relation, the initial conditions are fixed in this
region. The usual choice of the vacuum state is $B_1=1/\sqrt{2n}$,
$B_2=0$. In general $B_1$ and $B_2$ are determined by the matching
conditions between Regions 1 and 2. However, if the dynamics is
adiabatic throughout (in particular if the $a''/a$ term is
negligible), the WKB approximation holds and the solution is always
given by (\ref{WKBsol}). Therefore, if we start with $B_2 = 0$ and
uses this solution, one finds that $B_2$ remains zero at all
times. Deep in the region where $n_{\rm eff} \simeq n$ the solution
becomes
\begin{equation}
\mu (\eta ) \simeq {1 \over {\sqrt{2n}}} \exp(-i \phi - i n \eta),
\end{equation}
i.e. the standard vacuum solution times a phase which will disappear
when we calculate the modulus. The phase $\phi $ is given by $\phi
\equiv \int _{\eta _{\rm i}}^{\eta _1}n_{\rm eff} {\rm d}\tau $, where
$\eta _1$ is the time at which $n_{\rm eff}\simeq n$.
\par
The situation changes dramatically if we consider non-monotonic
dispersion relations. This is the case if $b_m<0$ in
Eq. (\ref{disp12}) or for the dispersion relation introduced in
Ref. \cite{Mersini}. Two new features can occur. Either the dispersion
relation can become complex (this is in general the case if $b_m<0$)
or the term $a''/a$ dominates (this is the case for the dispersion
relation introduced in Ref. \cite{Mersini} or for the Corley/Jacobson
dispersion relation with $b_m<0$ for a certain range of comoving
wavenumbers). In this case, the WKB approximation is violated in
Region 1 and we expect changes in the final spectrum. Unfortunately,
in the context of the standard increasing inflationary scale factor,
one also looses the ability to fix natural initial
conditions. However, this is no longer true if the spacetime is
asymptotically flat because then, at infinity, the term $a''/a$ goes
to zero, see also Ref.~\cite{LLMU}. Therefore, in this case, we can
choose well-motivated initial conditions. A good example of such a
situation is provided by a bouncing Universe. From the above
considerations, we expect that in this case the final spectrum is
modified and the initial conditions can be fixed naturally.  As an
additional benefit, complex frequencies can be avoided. In the next
section, we consider a toy model where these qualitative arguments can
be implemented concretely, at the level of equations.

\section{A Specific Example}

\begin{figure}[ht]
\includegraphics*[width=8.5cm, angle=0]{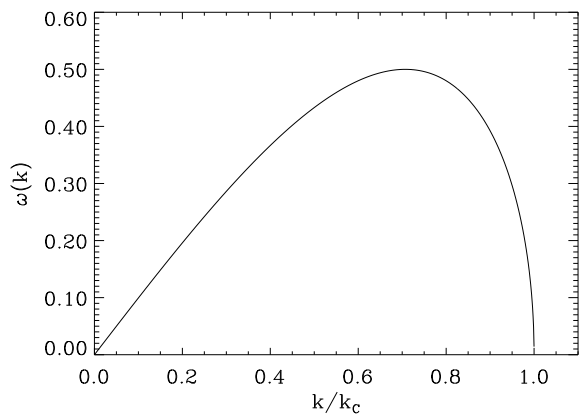}
\caption{Corley/Jacobson dispersion relation for $m=1$ and $\vert
b_m\vert =1$. In general $\omega _{_{\bf phys}}^2(k)$ vanishes at
$k=k_{_{\rm C}}\vert b_m \vert ^{-1/(2m)}$. For $k>k_{_{\rm C}}\vert
b_m \vert ^{-1/(2m)}$, the physical frequency becomes imaginary.}
\label{cjdisp}
\end{figure}

We will now illustrate the qualitative arguments of the previous
section with a concrete quantitative example. We take the
asymptotically flat bouncing Universe given by (\ref{model2}).  We
consider the type of modified dispersion relation for which in the
case of an expanding inflationary Universe the deviations in the
spectrum were found \cite{BM1,MB2}, namely the (generalization of the)
Corley/Jacobson dispersion relation (\ref{disp12}) with $b_m$
negative. In this case, $\omega _{_{\bf phys}}(k)$ is given by
\begin{equation}
\label{disp}
\omega _{_{\bf phys}}^2(k)=k^2-\vert b_m\vert 
k^2\biggl(\frac{k^2}{k_{_{\rm C}}^2}\biggr)^m,
\end{equation}
where $k_{_{\rm C}}$ is the cutoff physical wavenumber. The dispersion 
relation is represented in Fig.~\ref{cjdisp}.

\begin{figure}[ht]
\includegraphics*[width=8.5cm, angle=0]{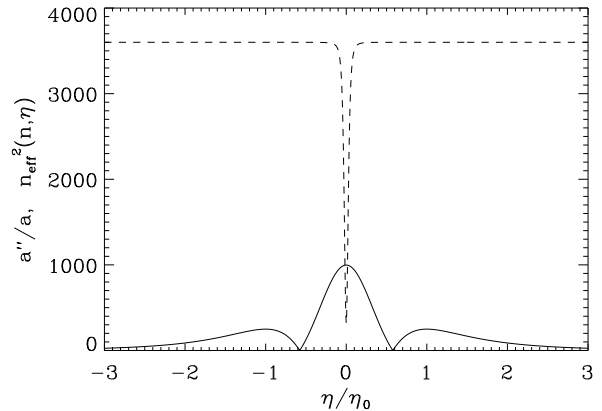}
\caption{The solid curve is $a''/a$, the dashed one $n^2_{\rm eff}$. The 
values used are $\ell _0=5000$, $\ell _{\rm b}=10$, $\ell _{_{\rm C}}=1$, 
$\vert b_m\vert =1$, $m=1$ and $n=60$.}
\label{neff}
\end{figure}

\begin{figure}[ht]
\includegraphics*[width=8.5cm, angle=0]{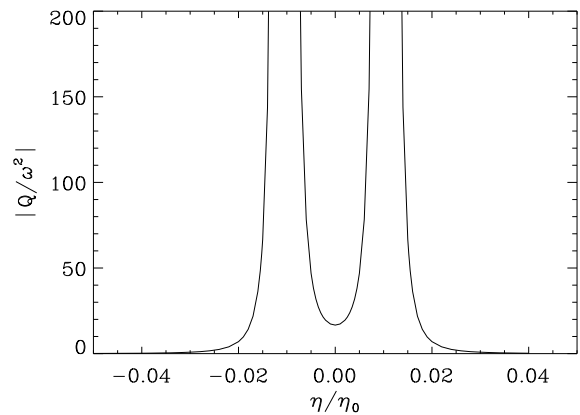}
\caption{Absolute value of Q versus conformal time. The 
values used are $\ell _0=5000$, $\ell _{\rm b}=10$, $\ell _{_{\rm C}}=1$, 
$\vert b_m\vert =1$, $m=1$ and $n=55$.}
\label{Qwkb}
\end{figure}

The behavior of the solutions of the equation of motion 
$\mu ''+\mu(n_{\rm eff}^2-a''/a)=0$ is determined by the competition 
between the two terms $n_{\rm eff}^2$ and $a''/a$ given by
\begin{widetext}
\begin{equation}
\label{wmu}
n_{\rm eff}^2(n,\eta) \, = \, n^2-n^2\frac{\vert b_m \vert }{\ell
_{_{\rm C}}^{-2m}} \biggl(\frac{n}{2\pi}\biggr)^{2m} \biggl[\ell
_0-\frac{\ell _0-\ell _{\rm b}}{1+( \eta /\eta _0)^2}\biggr]^{-2m},
\quad \frac{a''}{a}= \frac{2(\ell _0-\ell _{\rm b})}{\eta _0^2\ell
_{\rm b}} \frac{1-3(\eta /\eta_0)^2}{[1+(\eta /\eta _0)^2]^3},
\end{equation}
where $\ell _{_{\rm C}}\equiv 1/k_{_{\rm C}}$ is the cutoff
length. These two terms are represented in Fig.~\ref{neff}. When
$n_{\rm eff}^2>a''/a$, which is always the case when $\eta /\eta _0
\rightarrow \pm \infty$, the WKB approximation is valid and the
fundamental solutions can be written as
\begin{equation}
\label{wkb}
\mu \simeq \frac{1}{\sqrt{2n_{\rm eff}}}\exp \biggl
[\pm i\int ^{\eta }n_{\rm eff}(n,\tau ) {\rm d}\tau\biggr]
\rightarrow _{\eta /\eta _0\rightarrow \pm \infty}\frac{1}{\sqrt{2n}}
e^{\pm in\eta }.
\end{equation}
\end{widetext}
In this case, the physical wavenumbers of the modes are such that 
they correspond to the linear part of the dispersion 
relation, see Fig.~\ref{cjdisp}. In this situation, natural 
initial conditions can be chosen. Note that these initial conditions 
are, in a sense, even ``more standard'' than in the previous studies on 
trans-Planckian physics since usually the initial conditions 
are set in a region where the dispersion relation is not 
linear (but, as explained above, since the WKB approximation holds, 
meaningful initial conditions can nevertheless be considered). On the 
contrary, when $n_{\rm eff}^2<a''/a$ 
the WKB approximation is violated and two independent solutions are 
given by
\begin{equation}
\label{superh}
\mu \simeq a(\eta ), \quad 
\mu \simeq a(\eta )\int ^{\eta }\frac{{\rm d}\tau }{a^2({\tau })}.
\end{equation}
In this case, we are close to the cutoff scale and $\omega _{_{\bf
phys}}^2(k)$ being close to zero, see Fig.~\ref{cjdisp}, the $a''/a$
term dominates. This situation is very similar to the one discussed in
Ref. \cite{Mersini} where this part of the dispersion relation has
been named ``the tail''. In fact, it is possible to quantify exactly
the accuracy of the WKB approximation. Given an equation of the
form $\mu ''+\omega ^2\mu =0$, the WKB approximation is valid if the
quantity $\vert Q/\omega ^2 \vert \ll 1$, where the quantity $Q$ is
defined by the following expression
\begin{equation}
Q=\frac{3(\omega ')^2}{4\omega ^2}-\frac{\omega ''}{2\omega }.
\end{equation}
This standard criterion can be obtained in the following manner. The
WKB solution, $\mu _{\rm wkb}$, satisfies the equation $\mu _{\rm
wkb}''+\mu _{\rm wkb}(\omega ^2-Q)=0$ exactly. Therefore, one has $\mu
\simeq \mu _{\rm wkb}$ if $\vert Q/\omega ^2 \vert \ll 1$. In the
present context, $\omega ^2 $ is of course equal to $n_{\rm
eff}^2-a''/a$. We have plotted the quantity $\vert Q/\omega ^2\vert $
in Fig.~\ref{Qwkb}. When $\vert \eta /\eta _0 \vert \gg 1$, $\vert
Q/\omega ^2 \vert \ll 1$ and the WKB approximation is satisfied, as
expected from the previous discussion. The quantity $\vert Q/\omega
^2\vert $ increases when one approaches the bounce. At $\eta = \eta
_{\rm j}$, where $\omega (\eta _{\rm j})=0$, there is a
divergence. This point is a turning point. Then for $\vert \eta /\eta
_0 \vert \ll 1$, as can be seen in Fig.~\ref{Qwkb}, $\vert Q/\omega
^2\vert $ remains large and the WKB approximation breaks down in
agreement with the considerations above.

In a bouncing Universe, the modes start in the linear part of the
dispersion relation, pass through the non-linear part and then come
back in the standard region. However, things are not so simple because
there exists a range of comoving wavenumbers such that the dispersion
relation becomes complex. Although this case is {\it a priori}
interesting, it clearly requires more speculative considerations, in
particular the quantization in the presence of imaginary frequency
modes. Since our goal in this paper is to exhibit a case where
everything can be done in a standard manner, we will restrict
ourselves to modes which never enter the region where the dispersion
relation becomes complex. In the following, we determine the range of
comoving wavenumbers we are interested in and for which we are going
to calculate the power spectrum.  
\par 

The maximum of the absolute value of $a''/a$ is located at $\eta =0$
and is given by $(2/\eta _0^2)(\ell _0/\ell _{\rm b}-1)$.  Therefore,
if we restrict ourselves to modes such that
\begin{equation} 
\label{bound}
n \, \gg \, n_{\rm b}\equiv \frac{\sqrt{2}}{\eta _0}\biggl(
\frac{\ell _0}{\ell _{\rm b}}-1\biggr)^{1/2},
\end{equation}
then one is sure that, with an unmodified dispersion relation, the term 
$a''/a$ can always be neglected and that the initial spectrum is never 
changed. Clearly, this is not a physical restriction but it renders the 
comparison with the case with a modified dispersion relation easier. The 
minimum of the modified part of the dispersion relation is given 
by (from now on, we consider the case $\vert b_m\vert =1$ and $m=1$ 
since it is more convenient and does not restrict the physical 
content of the problem in any way)
\begin{equation}
n_{\rm eff}^2(n,\eta=0)=n^2
\biggl[1-\biggl(\frac{n}{n_{\rm inf}}\biggr)^2\biggr], \quad 
n_{\rm inf}\equiv 2\pi\frac{\ell _{\rm b}}{\ell
  _{_{\rm C}}}.
\end{equation}
Therefore, to maintain a real dispersion relation, we should only
consider modes such that $n < n_{\rm inf}$. Of course, for
consistency, the parameters of the model must be chosen such that
$n_{\rm b} < n_{\rm inf}$. This is the case for the most natural
choice, i.e., $\ell _0 \gg \ell _{\rm b} \gg \ell _{_{\rm C}}$. In the
following, as already mentioned above, the time $\eta _{\rm j}(n)$ such
that
\begin{equation}
n_{\rm eff}^2(n,\eta _{\rm j}) \, = \, {{a''} \over a}
\end{equation}
will play a crucial role.  For convenience we will only consider
values of $n$ such that this time is determined only by the central
peak of $a''/a$ and not by the two wings. In practice, this amounts to
taking $n_{\rm eff}^2(n,\infty ) > n_{\rm b}^2/4$ since $n_{\rm
b}^2/4$ is the maximum height of the wings. Then it is easy to show
that $n^2\in [n_-^2,n_+^2]$ where
\begin{equation}
n_{\pm }^2=\frac{n_{\rm sup}^2}{2}\
\biggl(1\pm \sqrt{1-\frac{n_{\rm b}^2}{n_{\rm sup}^2}}\biggr),
\end{equation}
where $n_{\rm sup}\equiv 2\pi \ell _0/\ell _{_{\rm C}}$ is the 
largest value of $n$ such that $n_{\rm eff}^2(n,\infty )$ remains 
real. Therefore, the range we are interested in is given by
\begin{equation}
\max (n_{\rm b}^2,n_-^2)<n^2<\min(n_{\rm inf}^2,n_+^2).
\end{equation}
In practice, we have $n_-^2 \simeq (1/4)n_{\rm b}^2$ and 
$n_+^2\simeq n_{\rm sup}^2$. This means that 
$\max(n_{\rm b}^2,n_-^2)=n_{\rm b}^2$, 
$\min(n_{\rm inf}^2,n_+^2)=n_{\rm inf}^2$ and that 
the range reduces to 
\begin{equation}
n_{\rm b}^2 \,< n^2 \, < \, n_{\rm inf}^2 \, .
\end{equation}
\par
\begin{figure}[ht]
\includegraphics*[width=8.5cm, angle=0]{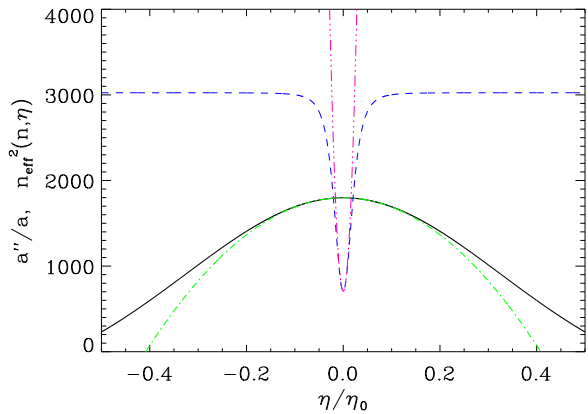}
\caption{The solid curve represents $a''/a$, the dot-dashed curve its
quadratic approximation. The dashed curve represents $n^2_{\rm eff}$,
and its quadratic approximation is the fourth curve. Values chosen are
$\ell _0=5000$, $\ell _{\rm b}=10$, $\ell _{_{\rm C}}=1$, $\vert b_m\vert
=1$, $m=1$ and $n=60$.}
\label{approx}
\end{figure}
\begin{figure}[ht]
\includegraphics*[width=8.5cm, angle=0]{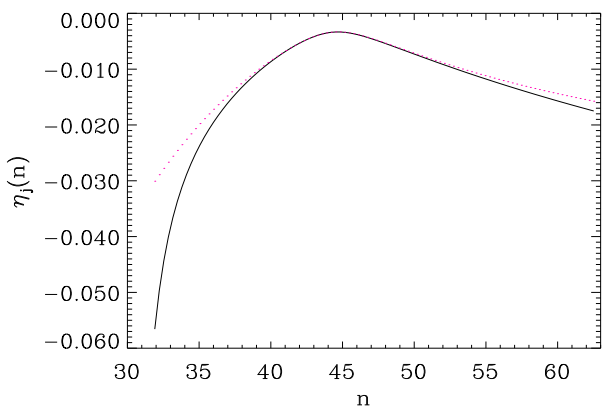}
\caption{The dependence of the matching time $\eta_{\rm j}(n)$ on the
wave number $n$ for the values $\ell _0=5000$, $\ell _{\rm b}=10$,
$\ell _{_{\rm C}}=1$, $\vert b_m\vert =1$, $m=1$. The solid curve is the
exact result determined numerically, the dotted curve is obtained
using the analytical approximations.}
\label{etaj}
\end{figure}

Having determined the relevant wavenumbers, we can now choose the 
initial condition and solve the equation of motion. In the 
first region where $\eta <-\eta _{\rm j}(n)$, we only consider positive 
frequency modes and we have
\begin{equation} 
\mu _{\rm I}(\eta )=\frac{1}{\sqrt{2n_{\rm eff}}}\exp \biggl
[-i\int ^{\eta }_{\eta _{\rm i}}n_{\rm eff}(n,\tau ) 
{\rm d}\tau \biggr],
\end{equation}
where $\eta _{\rm i}$ is an arbitrary initial time. In the 
second region, where $-\eta _{\rm j}(n)<\eta <\eta _{\rm j}(n)$, the 
solution is given by
\begin{equation}
\mu _{\rm II}(\eta )= B_1 a(\eta )+B_2a(\eta )\int _0^{\eta }
\frac{{\rm d}\tau }{a^2({\tau })}.
\end{equation}
The lower bound of the integral is {\it a priori} arbitrary. However,
it is very convenient to take it equal to zero because in this case
the second branch becomes odd whereas the first one (i.e. the scale
factor) is even. Then, it is easy to show that
\begin{widetext}
\begin{equation}
\int _0^{\eta }\frac{{\rm d}\tau }{a^2({\tau })}= \frac{\eta _0}{\ell
_{\rm b}^2}\biggl\{ \frac{1}{p^2}\frac{\eta }{\eta _0}
+\frac{1-2p+p^2}{2p^2[1+p(\eta /\eta _0)^2]}\frac{\eta }{\eta _0}
+\frac{p^2+2p-3}{2p^{5/2}}\arctan\biggl(\sqrt{p}\frac{\eta }{\eta
_0}\biggr) \biggr\},
\end{equation}
where $p\equiv \ell _0/\ell _{\rm b}$. Finally, the solution in the 
third region where $\eta >\eta _{\rm j}(n)$ can be written as 
\begin{equation} 
\mu _{\rm III}(\eta )=\frac{C_1}{\sqrt{2n_{\rm eff}}}\exp \biggl
[-i\int ^{\eta }_{\eta _{\rm i}}n_{\rm eff}(n,\tau ) 
{\rm d}\tau\biggr]+\frac{C_2}{\sqrt{2n_{\rm eff}}}\exp \biggl
[+i\int ^{\eta }_{\eta _{\rm i}}n_{\rm eff}(n,\tau ) 
{\rm d}\tau\biggr].
\end{equation}
The goal is now to calculate the coefficients $C_1$ and $C_2$. Using 
the continuity of the mode function $\mu $ and of its derivative, 
we find
\begin{eqnarray}
\label{coeffC1}
C_1(n) &=& \frac{i}{2W[-\eta _j(n)]}\frac{e^{i[\Omega (\eta _{\rm j})
-\Omega (-\eta _{\rm j})]}}{\sqrt{n_{\rm eff}(n,-\eta _{\rm j}) 
n_{\rm eff}(n,\eta _{\rm j})}}\biggl\{-[g'+\alpha g](-\eta _{\rm j})
[f'+\bar{\alpha }f](\eta _{\rm j})+[f'+\alpha f](-\eta _{\rm j})
[g'+\bar{\alpha }g](\eta _{\rm j})\biggr\}, \\
\label{coeffC2}
C_2(n) &=&
\frac{-i}{2W[-\eta _j(n)]}\frac{e^{-i[\Omega (\eta _{\rm j})
+\Omega (-\eta _{\rm j})]}}{\sqrt{n_{\rm eff}(n,-\eta _{\rm j}) 
n_{\rm eff}(n,\eta _{\rm j})}}\biggl\{-[g'+\alpha g](-\eta _{\rm j})
[f'+\alpha f](\eta _{\rm j})+[f'+\alpha f](-\eta _{\rm j})
[g'+\alpha g](\eta _{\rm j})\biggr\}, 
\end{eqnarray}
where we have used the short-hand notation $f\equiv a(\eta )$ and 
$g\equiv a(\eta )\int ^{\eta }_0{\rm d}\tau /a^2(\tau )$. We 
have also utilized the following definitions: $W\equiv gf'-g'f$ is the 
Wronskian, $\Omega (\eta )\equiv 
\int ^{\eta }_{\eta _{\rm i}}n_{\rm eff}(n,\tau ) {\rm d}\tau $, and 
the quantity $\alpha $ is 
\begin{equation}
\label{defalpha}
\alpha \equiv \frac{n_{\rm eff}'}{2n_{\rm eff}}+in_{\rm eff}.
\end{equation}
Using the fact that $n_{\rm eff}$ is even, it is easy to see that
$\alpha (n,-\eta )=-\bar{\alpha}(n,\eta )$. Using the same property,
one could also simplify the factor $\sqrt{n_{\rm eff}(n,-\eta _{\rm
j}) n_{\rm eff}(n,\eta _{\rm j})}=n_{\rm eff}(n,\eta _{\rm j})$ in the
above equation. The final result is given by Eqs.~(\ref{coeffC1}),
(\ref{coeffC2}) where all the functions are explicitly known except
the function $\eta _{\rm j}=\eta _{\rm j}(n)$. The dependence on the
dispersion relation of the final result is completely encoded in this
function. The time $\eta _{\rm j}(n)$ is the solution of an algebraic
equation that is not possible to solve explicitly.  However, we can
find an approximation if the function $n_{\rm eff}(n,\eta )$ is Taylor
expanded around $\eta =0$. For the function $a''/a$ , we use a
quadratic least square approximation which gives a better result. We
obtain
\begin{equation}
\label{Taexpand}
n_{\rm eff}^2(n,\eta )\simeq n^2-n^2\biggl(\frac{n}{n_{\rm inf}}\biggr)^2
(1-n_{\rm b}^2\eta ^2), \quad \frac{a''}{a} \simeq n_{\rm b}^2\biggl[
1-6\biggl(\frac{\eta }{\eta _0}\biggr)^2\biggr].
\end{equation}
The two functions and their approximations are represented in
Fig.~\ref{approx}. We see now that $\eta _{\rm j}(n)$ is the
solution of a quadratic algebraic equation. Straightforward
calculations give
\begin{equation}
\label{approxetaj}
\eta _{\rm j}^2(n) \simeq \frac{n^2}{n_{\rm b}^2}
\biggl[\biggl(\frac{n}{n_{\rm inf}}\biggr)^2+
\biggl(\frac{n_{\rm b}}{n}\biggr)^2-1\biggr]
\biggl[n^2\biggl(\frac{n}{n_{\rm inf}}\biggr)^2+
\frac{6}{\eta _0^2}\biggr]^{-1}.
\end{equation}
\end{widetext}
The function $\eta _{\rm j}(n)$ can also be found exactly by numerical
calculations. 
We have done this computation by means of a Fortran
code. The comparison of the exact result with its approximation given
in Eq. (\ref{approxetaj}) is represented in Fig.~\ref{etaj}. The
function $n_{\rm eff}(n,\eta =0)$ is not monotonic and has a maximum
around $n\simeq 45$ for the values of the parameters considered in
Fig.~\ref{etaj}. For this value $n_{\rm eff}(n,\eta )$ is almost
never smaller than $a''/a$. It is clear that around this value the
approximation will be very good whereas for other values of $n$,
especially for $n\simeq n_{\rm b}$, the approximation will be less
good. This is due to the fact that when $n$ approaches $n_{\rm b}$,
the asymptotic value of $n_{\rm eff}(n,\eta )$ decreases and the curve
opens out at the intersection with $a''/a$. As a consequence, the
quadratic approximation employed above breaks down. This is confirmed
by the plots in Fig.~\ref{etaj}. The approximation can be less good
for other values of the parameters.
\par
From the above analysis, it is clear that $\epsilon \equiv \eta _{\rm
j}(n)/\eta _0$ is a small number. We can therefore expand everything 
in terms of the parameter $\epsilon$. This will allow us to get 
an analytical estimate of the spectrum. After tedious but 
straightforward computations, one obtains the following expressions 
for the coefficients $C_1(n)$ and $C_2(n)$ 
\begin{widetext}
\begin{eqnarray}
C_1(n) &=& 1+\epsilon \frac{i\eta _0}{2n}\biggl[
2n^2\biggl(1-\frac{n^2}{n_{\rm inf}^2}\biggr)^{1/2} +n_{\rm
b}^2\biggl(1-\frac{n^2}{n_{\rm inf}^2}\biggr)^{-1/2} +n_{\rm
b}^2\biggl(1-\frac{n^2}{n_{\rm inf}^2}\biggr)^{-3/2}\biggr] 
+{\cal O}(\epsilon ^2),
\label{c1c2a}
\\ 
C_2(n)&=& -\epsilon \frac{i\eta _0}{2n} 
e^{-2i\Psi _{\rm i}(n)}
\biggl[
2n^2\biggl(1-\frac{n^2}{n_{\rm inf}^2}\biggr)^{1/2} +n_{\rm
b}^2\biggl(1-\frac{n^2}{n_{\rm inf}^2}\biggr)^{-1/2} +n_{\rm
b}^2\biggl(1-\frac{n^2}{n_{\rm inf}^2}\biggr)^{-3/2}\biggr]
+{\cal O}(\epsilon ^2),
\label{c1c2b}
\end{eqnarray}
\end{widetext}
where $\Psi _{\rm i}(n)\equiv \int _{\eta _{\rm i}}^0n_{\rm
eff}(n,\tau ){\rm d}\tau $.  Several comments are in order at this
point. Firstly, the expressions of the two coefficients have exactly
the expected form. When the parameter $\epsilon $ goes to zero,
$C_1(n)$ goes to one and $C_2(n)$ goes to zero, i.e., we recover an
unmodified spectrum. This is in complete agreement with the fact that,
when $\epsilon = \eta _{\rm j}(n)/\eta _0 \rightarrow 0$, the
effective wavenumber $n_{\rm eff}$ never penetrates the region where
the WKB approximation is violated. This is in this sense that the
parameter $\epsilon $ is the relevant parameter to expand the spectrum
in since it is clearly directly linked to the violation of the WKB
approximation. Therefore, in the limit $\epsilon \rightarrow 0$, one
recovers the usual spectrum. Secondly, it is interesting to notice
that the two terms proportional to $\epsilon $ in $C_1(n)$ and
$C_2(n)$ are the same up to a global phase. To compute the 
phase in Eqs.~(\ref{coeffC1}) and (\ref{coeffC2}), we have 
written $\Omega (\eta )=\Psi _{\rm i}(n)+\int _{0}^{\eta }n_{\rm
eff}(n,\tau ){\rm d}\tau $. Since we only need to evaluate 
$\Omega (\pm \eta _{\rm j})$, one can use the approximate equation 
for $n_{\rm eff}$, see Eq.~(\ref{Taexpand}), in the integral 
$\int _{0}^{\eta _{\rm j}}n_{\rm eff}(n,\tau ){\rm d}\tau $. One 
finds 
\begin{equation}
\Omega (\pm \eta _{\rm j})=\Psi _{\rm i}\pm \epsilon 
n\eta _0\biggl(1-\frac{n^2}{n_{\rm inf}^2}\biggr)^{1/2}+{\cal O}(\epsilon ^3).
\end{equation}
Let us remark that the term $e^{-i[\Omega (\eta _{\rm j})+\Omega
(-\eta _{\rm j})]}$ in the expression of $C_2(n)$ does not contribute
to order $\epsilon $ whereas the factor $e^{i[\Omega (\eta _{\rm
j})-\Omega (-\eta _{\rm j})]}$ gives a contribution at this
order. This contribution is exactly such that the two terms
proportional to $\epsilon $ in Eqs.~(\ref{c1c2a}), (\ref{c1c2b}) have
the same absolute value. Thirdly, the factor $1-n^2/n_{\rm inf}^2$ is
always positive since we restrict ourselves to the regime where
$n^2<n_{\rm inf}^2$. Therefore, the above expressions of $C_1(n)$ and
$C_2(n)$ are always well-defined. Finally, interestingly enough, the
two coefficients $C_1(n)$ and $C_2(n)$ have a similar structure to 
one of the corresponding coefficients found in Ref.~\cite{LLMU} where
another example of a modified spectrum has been exhibited.  \par
\begin{figure}[ht]
\includegraphics*[width=8.5cm, angle=0]{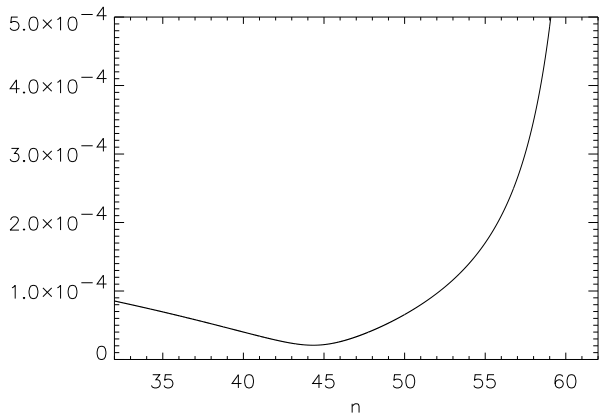}
\caption{Amplitude of the next-to-leading order correction to the
spectrum without the superimposed oscillations. Values chosen are
$\ell _0=5000$, $\ell _{\rm b}=10$, $\ell _{_{\rm C}}=1$, $\vert b_m\vert
=1$, $m=1$.  When $n$ approaches $n_{\rm inf}$ the correction blows up
and the approximation becomes meaningless.}
\label{ampli}
\end{figure}

 We
are now in a position where we can calculate the spectrum
explicitly. It is defined by the following equation
\begin{equation}
n^3P(n)\propto n^3\biggl \vert \frac{\mu }{a} \biggr \vert ^2.
\end{equation}
We will evaluate the spectrum in the region where $\eta /\eta _0\gg
1$. In this case, $a(\eta )\simeq \ell _0$ and $n_{\rm eff}\simeq
n$. Expanding everything in terms of the small parameter $\epsilon $,
one obtains
\begin{widetext}
\begin{eqnarray}
n^3P(n)&\propto & \frac{n^2}{2\ell _0^2}+\epsilon \frac{\eta _0}{2\ell _0^2}
n\biggl[
2n^2\biggl(1-\frac{n^2}{n_{\rm inf}^2}\biggr)^{1/2} +n_{\rm
b}^2\biggl(1-\frac{n^2}{n_{\rm inf}^2}\biggr)^{-1/2} +n_{\rm
b}^2\biggl(1-\frac{n^2}{n_{\rm inf}^2}\biggr)^{-3/2}\biggr]
\nonumber \\
& &\times 
\sin \biggl[2\Phi (n)
-2\Psi _{\rm i}(n)+2n(\eta -\eta _+)\biggr] +{\cal O}(\epsilon ^2),
\end{eqnarray}
\end{widetext}
where $\eta _+>\eta _{\rm j}$ is the time at which $n_{\rm eff} \simeq
n$ and the phase $\Phi (n)$ is defined by $\Phi (n)\equiv \int _{\eta _{\rm
i}}^{\eta _+}{\rm d}\tau n_{\rm eff}$. Again the spectrum has the
expected form. The leading order is the vacuum spectrum with spectral
index $3$ and the next-to-leading order (proportional to $\epsilon $)
has the form of a complicated function of $n$ times superimposed
oscillations. We have plotted this complicated function (without the
superimposed oscillations) in Fig.~\ref{ampli}.
We can check that the correction remains small and that, therefore,
the approximation employed in this article is consistent. On the other
hand, when the wavenumber $n$ approaches $n_{\rm inf}$, the
approximation scheme breaks down but, precisely in this limit, the
whole problem becomes pointless in agreement with the considerations
above. We also see that the deviation is minimum around $n\simeq 45$
(for these values of the parameters) because, as already mentioned
above, in this case the term $a''/a$ almost never dominates. Although
the present example is only a toy model, it is quite interesting to
see that it bears some close resemblance with the other example of a
modified spectrum found in Ref.~\cite{LLMU}. This suggests that the
features previously exhibited may be valid in general.  
\par 
We have thus reached our main goal: find an example where the initial
conditions can be fixed naturally, where the frequency never becomes
complex and where the final spectrum is modified. Note that in this
example, the change in the spectrum comes about from an interplay
between the modified dispersion relation factor $n^2_{\rm eff}$ and
the factor $a''/a$ which is responsible for Parker particle production
\cite{Parker}. For an unmodified linear dispersion relation, the
$a''/a$ term is always negligible and there is no particle production.
However, for our modified dispersion relation, for a range of modes
the Parker particle production term becomes important and leads to
non-adiabatic evolution. The length of time during which the $a''/a$
term dominates depends on the specifics of the dispersion relation,
and hence the final spectrum will depend on these specifics.
 
\section{Discussion}

In this paper we have studied the dependence of the spectrum of a
free scalar field in a bouncing Universe on trans-Planckian effects 
introduced via a modification of the free field dispersion relation.
We have found that both the amplitude and the slope of the spectrum
depend on the dispersion relation, assuming that the dispersion
relation leads to non-adiabatic evolution of the mode functions
on trans-Planckian physical length scales. Such non-adiabatic
evolution is possible without requiring the effective frequency to
become imaginary.
\par
This result supports our earlier work which shows that the spectrum of
cosmological fluctuations in inflationary cosmology can depend
sensitively on trans-Planckian physics \cite{BM1,MB2}. Note that
results supporting \cite{BM1,MB2} were recently also obtained by
\cite{EGKS} (see also Ref.~\cite{KN}) in the context of a mode
equation modified from the usual ones by taking into account
\cite{Kempf} effects coming from a short distance cutoff. For a recent
paper on the effects of short-distance physics on the consistency
relation for scalar and tensor fluctuations in inflationary cosmology
see Ref.~\cite{Hui}.  The calculation we presented here is free of two
of the possible objections against the earlier work. The first
objection was that the use of the minimum energy density state as
initial state is not well defined on trans-Planckian scales if the
dispersion relation differs dramatically from the linear one (which it
has to in order to get non-adiabatic evolution). In our present work,
the initial conditions are set in the low curvature region and on
length scales larger than the cutoff length but smaller than the
Hubble radius, where the choice of initial vacuum state is well
defined.  The second objection concerned the use of dispersion
relations for which the effective frequency is imaginary in some time
interval. No imaginary frequencies are used in the present model.
\par
The results obtained in this paper apply also both to bouncing
Universe backgrounds described by the two other models mentioned 
in the Introduction. In the case of the first model (where 
the scale factor is given by an hyperbolic cosine), similar
results would hold if we would match the {\it local Minkowski
vacuum} (WKB vacuum) at some initial time and express the
results in terms of the WKB vacuum state at the corresponding
post-bounce time. However, in that case, the
initial conditions are not easy to justify since they have to
be set on length scales larger than the Hubble radius. Obviously,
we could consider that model and let the scale
factor make a further transition to an asymptotically radiation-
dominated or asymptotically flat stage at very large initial
and final times. In this case, it is very likely that results
similar to the ones we obtained here in model (\ref{model2})
would be obtained.

\vspace{0.5cm} 
\centerline{\bf Acknowledgments}
\vspace{0.2cm}

We would like to thank M.~Lemoine for comments and careful reading 
of the manuscript. We acknowledge support 
from the BROWN-CNRS University Accord which made possible the visit of 
J.~M. to Brown during which most of the work on this project was done, 
and we are 
grateful to Herb Fried for his efforts to secure this Accord. S.~E.~J. 
acknowledges financial support from CNPq. R.~B. 
wishes to thank Bill Unruh for hospitality at the University of British 
Columbia during the time when this work was completed, and for many
stimulating discussions. J.~M. thanks the High Energy 
Group of Brown University for warm hospitality. The research was supported in 
part by the U.S. Department of Energy under Contract DE-FG02-91ER40688, TASK A.

\end{document}